\documentclass[conference]{IEEEtran}
\IEEEoverridecommandlockouts
\usepackage{cite}
\usepackage{graphicx}
\usepackage{textcomp}
\usepackage{xcolor}
\usepackage{multirow}
\usepackage{booktabs} % usage of multicolumns in tables
\usepackage{hhline}  % Add the hhline package
\usepackage{hyperref}
\usepackage{subcaption}
\usepackage[inline]{enumitem}
\usepackage{ulem}

\newcommand{\MyFormat}{CAPG}

\usepackage{todonotes}
\usepackage{lipsum}
\usepackage{tabularx}
\usepackage{makecell}

\usepackage{listings}
\usepackage{float} % to include \begin{figure}[H]
\usepackage{color}

\definecolor{teal}{rgb}{0.0, 0.5, 0.5}
\definecolor{blush}{rgb}{0.87, 0.36, 0.51}
\definecolor{applegreen}{rgb}{0.55, 0.71, 0.0}
\definecolor{amethyst}{rgb}{0.6, 0.4, 0.8}
\definecolor{ultramarine}{RGB}{0,32,96}
\definecolor{delim}{RGB}{20,105,176}
\definecolor{numb}{RGB}{106, 109, 32}
\definecolor{string}{rgb}{0.64,0.08,0.08}
\definecolor{key-color}{rgb}{0.8, 0.47, 0.196}

\lstdefinelanguage{json}{
    frame=single,
    rulecolor=\color{black},
    showspaces=false,
    showtabs=false,
    breaklines=true,
    breakatwhitespace=true,
    basicstyle=\ttfamily\small,
    upquote=true,
    morestring=[b]",
    stringstyle=\color{string},
    otherkeywords = {type, name, id},
    morekeywords = {type, name, id},
    keywordstyle = {\color{key-color}},
    literate=
    *{0}{{{\color{numb}0}}}{1}
        {1}{{{\color{numb}1}}}{1}
        {2}{{{\color{numb}2}}}{1}
        {3}{{{\color{numb}3}}}{1}
        {4}{{{\color{numb}4}}}{1}
        {5}{{{\color{numb}5}}}{1}
        {6}{{{\color{numb}6}}}{1}
        {7}{{{\color{numb}7}}}{1}
        {8}{{{\color{numb}8}}}{1}
        {9}{{{\color{numb}9}}}{1}
        {\{}{{{\color{delim}{\{}}}}{1}
        {\}}{{{\color{delim}{\}}}}}{1}
        {[}{{{\color{delim}{[}}}}{1}
        {]}{{{\color{delim}{]}}}}{1},
}
\usepackage{flushend}

\def\BibTeX{{\rm B\kern-.05em{\sc i\kern-.025em b}\kern-.08em
T\kern-.1667em\lower.7ex\hbox{E}\kern-.125emX}}

\begin{document}

    \title{
    %  UVDF: the Ultimate Vulnerability Description Format
        CVE representation to build attack positions graphs
    }

    \author{\IEEEauthorblockN{Manuel Poisson}
    \IEEEauthorblockA{%\textit{dept. name of organization (of Aff.)} \\
        \textit{AMOSSYS, CentraleSupélec, CNRS, Inria, Univ. Rennes, IRISA}\\
% TODO City, Country \\
        manuel.poisson@irisa.fr
    }
    \and
    \IEEEauthorblockN{Valérie Viet Triem Tong}
    \IEEEauthorblockA{%\textit{dept. name of organization (of Aff.)} \\
        \textit{CentraleSupelec, CNRS, Inria, Univ. Rennes, IRISA}\\
% TODO City, Country \\
        valerie.viettriemtong@centralesupelec.fr
    }
    \and
    \IEEEauthorblockN{Gilles Guette}
    \IEEEauthorblockA{%\textit{dept. name of organization (of Aff.)} \\
        \textit{Univ. Rennes, CNRS, Inria, IRISA}\\
% TODO City, Country \\
        gilles.guette@univ-rennes.fr
    }
    \and
%
%\IEEEauthorblockN{Erwan Abgrall }
%\IEEEauthorblockA{%\textit{dept. name of organization (of Aff.)} \\
%\textit{CentraleSupélec, Inria}\\
%erwan.abgrall@irisa.fr
%}
    \and
    \IEEEauthorblockN{Frédéric Guihéry}
    \IEEEauthorblockA{%\textit{dept. name of organization (of Aff.)} \\
        \textit{AMOSSYS}\\
% TODO City, Country \\
        frederic.guihery@amossys.fr
    }
    \and
    \IEEEauthorblockN{Damien Crémilleux}
    \IEEEauthorblockA{%\textit{dept. name of organization (of Aff.)} \\
        \textit{AMOSSYS}\\
% TODO City, Country \\
        damien.cremilleux@amossys.fr
    }
% TODO City, Country \\
% TODO email address or ORCID

    }

    \maketitle

    \begin{abstract}
        In cybersecurity, CVEs (Common Vulnerabilities and Exposures) are publicly disclosed hardware or software vulnerabilities. These vulnerabilities are documented and listed in the NVD database maintained by the NIST. Knowledge of the CVEs impacting an information system provides a measure of its level of security. This article points out that these vulnerabilities should be described in greater detail to understand how they could be chained together in a complete attack scenario.
        This article presents the first proposal for the  \MyFormat{} format, which is a method for representing a CVE vulnerability, a corresponding exploit, and associated attack positions.
    \end{abstract}


    \section{Introduction}
%The security level of an infrastructure can be assessed through audits or pentest campaigns, which reveal system vulnerabilities.
%These vulnerabilities are then generally described in the CVE format.
%This format structures the description of a vulnerability and mainly describes its nature, the hardware or software component on which it is found, and the purpose for which the vulnerability may be exploited.
%The severity of the vulnerability can be assessed by the CVSS score. This score is then translated into a qualitative representation (low, medium, high and critical) of the vulnerability, helping organizations to evaluate their risk management processes and prioritize the remediation.

Performing an audit of an information system (IS), such as an architecture audit, configuration audit or penetration test, are common approaches for mapping vulnerabilities.
These vulnerabilities can be exploited by attackers to increase their control over the system and propagate towards their ultimate objectives such as for instance, data exfiltration or ransomware execution.
Unfortunately, the estimation of the risk posed by combinations of vulnerabilities relies on auditors, who may overlook some attack paths. We argue here that the description of discovered vulnerabilities, in particular by CVEs and the various related formats, lacks precision to allow the automatic integration of their exploitation into an accurate and operational attack scenarios design. This paper proposes \MyFormat{} (\textit{from CVE to Attack Positions Graph}), a new format designed to represent exploits of CVEs and then  to highlight how multiple CVEs could be chained by attackers to spread themselves. The main purpose of \MyFormat{} is to highlight critical attack paths that need to be urgently monitored and corrected.

%It leverages the concept of attack positions, representing the ability to execute commands on a given machine with the rights of a specific user. \MyFormat{} is based on the representation of attack scenarios through a graph of attack positions~\cite{poisson_aware} highlighting attackers' propagation through the use of attack procedures.\MyFormat{} provides a mean to clarify the nature of vulnerabilities and thus automate the construction of such an attack graph.
%The interest in this format is twofold. First, it allows elucidation of which types of user accounts and which machines attackers must control to exploit a given CVE. Second, it clarifies how CVEs help attackers advance toward their target after a successful exploitation.
This article presents two main contributions. The first contribution is a new \MyFormat{} format allowing to represent exploits of CVEs. \MyFormat{} gives a precise representation of how a vulnerability can be exploited by an attacker, and more importantly, what is the gain for an attacker who has successfully exploited a given CVE.
The second contribution is a methodology to populate this format by executing an exploit applicable to the CVE described. Section~\ref{sec::relatedWork} presents the background and the related work and motivates the use of this new format. Section~\ref{sec:definition} specifies \MyFormat{} and Section~\ref{sec:fillRunningExploits} explains how to fill it, and presents an example of its use in building an attack scenario.

    %. \newpage
%. \newpage

\section{Background and related work}
\label{sec::relatedWork}

\subsection{Series of formats to deal with vulnerabilities}

\begin{figure*}[!htpb]
    \centering
    \includegraphics[width=0.70\textwidth]{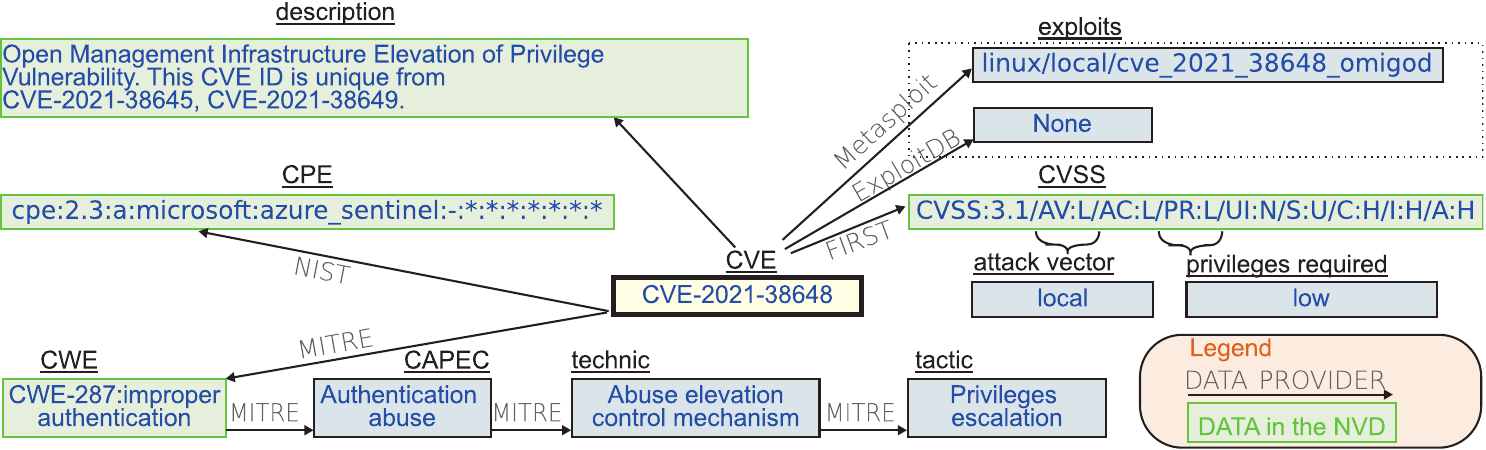}
    \caption{\texttt{CVE-2021-38648} and its related data}
    \label{fig:cve-and-data}
\end{figure*}

Common Vulnerabilities and Exposures (CVEs) are vulnerabilities that are publicly disclosed to protect against their exploitation.
All CVEs are indexed in the (US) national vulnerability database (NVD)~\cite{nvd_database}.
Penetration testers and attackers usually start by performing a reconnaissance of the information system they intend to infiltrate.
This phase includes a vulnerability scan where tools like Nessus~\cite{nessus} or OpenVAS~\cite{openvas} can reveal the presence of well-known vulnerabilities identified by a CVE-Id.
Many tools and knowledge bases use the CVE-Ids to uniquely identify a CVE.
For instance, Nessus can indicate if known exploits applicable to a given CVE are available.
It can be used to search for additional information on identified vulnerabilities.
The next step typically consists of gathering data related to these CVEs to understand and assess the associated risks.
In the NVD, a CVE is described by a free-form text description, and some additional resources depicted below.
As an example, for CVE 2021-38648 the following information is included as shown in Figure~\ref{fig:cve-and-data}.

\paragraph*{Severity of a CVE with the CVSS}
The Common Vulnerability Scoring System (CVSS), maintained by FIRST~\cite{cvss_spec_first}, intends to help prioritize responses and remediation.
The CVSS score is computed over a CVSS vector depicting how the CVE may be exploited, its ease of exploitation, and its impact on the target.
%A severity is assigned to each CVE depending on a score that follows the common vulnerability scoring system (CVSS) and that is computed based on its CVSS vector [X].The latter provides structured data about some characteristics of the CVE specified by the FIRST, a global organization working on incident response.For example the level of privileges an attacker must possess before successfully exploiting the vulnerability is given by the value ``None'', ``Low'' or ``High'' of the field ``Privileged Required''.Another CVE characteristic evaluated in the CVSS vector is the \textit{Attack vector} which, when having the value ``\textit{Local}'' means that ``\textit{the attacker exploits the vulnerability by accessing the target system locally}''or he ``\textit{relies on User Interaction by another person}''.

\paragraph*{Platforms vulnerable to a CVE with CPEs}
Each CVE is linked to a list of products that are known to be affected by the CVE.
These products are identified using Common Platform Enumerations (CPEs)~\cite{cpe}, a structured naming scheme for systems, hardware, and applications.
A CPE lists data like the type, the vendor, and the version of a product.

\paragraph*{Typical weaknesses responsible for the CVE}
In the NVD, CVEs are mapped to Common Weakness Enumerations (CWEs)~\cite{cwe} which define categories of weaknesses. CWEs are classified among over 900 types ranging from buffer overflows to cross-site scripting including hard-coded credentials, and insecure random numbers generators.

Data found in the NVD enables to estimate the security level of a system~\cite{williams_analyzing_2018} but does not allow understanding of how the discovered vulnerabilities can be exploited in a complete attack scenario. We argue that highlighting a full attack scenario based on the discovery of vulnerabilities requires precise information indicating both the pre-conditions for exploiting the vulnerability and the foothold attackers will have if they succeed in exploiting it.
In this context, the most useful information is given in the CVSS vector. The CVSS vector of \texttt{CVE-2021-38648} carries the information that the privileges required (\textit{PR}) to exploit this CVE are \textit{Low} (\textit{L}) and that the attack vector (\textit{AV}) is \textit{Local} (\textit{L}).
It means that there are no special conditions for exploiting the vulnerability except the ability to run commands directly on the vulnerable system (e.g.\ through physical access or SSH) and control over a local account.
This data is too imprecise to be integrated into an operational scenario.
The open data providing the most operational information is the potential exploit linked to the CVE. An exploit is a piece of code implemented to take advantage of a vulnerability.
Metasploit~\cite{noauthor_metasploit_nodate}, ExploitDB~\cite{exploitDB}, and GitHub~\cite{github} are three major public exploit sources~\cite{exploit_availability}.
% They are compared in Section~\ref{sec:fillRunningExploits} concerning their interest in gaining knowledge about CVEs.
However, limitations are that not all exploits are public and public exploits are not always related to a CVE and vice-versa.
%For example, one exploit is related to \href{https://nvd.nist.gov/vuln/detail/CVE-2021-38648}{\texttt{CVE-2021-38648}} in Metasploit but none in ExploitDB.
A deep analysis of an exploit should provide, the conditions that attackers have to fill to use it.
Unfortunately, the source code of exploits follows no particular specification and their metadata is not always consistent.
%In Metasploit, both exploits related to the CVEs \texttt{2021-38648} and \texttt{2022-39952} claim to lead to a session with the user \texttt{root}.
%However, their metadata indicating if the user reached after full execution of the exploit is ``\textit{Privileged}'' or not is set to ``\textit{No}'' and ``\textit{Yes}'' respectively
For example, in Metasploit, both exploits related to the CVEs \texttt{2021-38648} and \texttt{2022-39952}
lead to a session with the user \texttt{root} following their documentation, but
their metadata indicating if the user reached after their full execution is ``\textit{Privileged}'' or not is respectively set to ``\textit{No}'' and ``\textit{Yes}''.
Also, the source code of exploits is not enough to automatically derive neither the context in which they can be used nor the gain from their successful execution.

The lack of an appropriate format makes it very hard to find comprehensive data to derive pre/post conditions of a CVE exploitation.

\subsection{Related work}

%\manu{reference aux positions d'attaque}

Some research papers and tools intend to make sense of CVEs and their related data.
BRON~\cite{hemberg_bron} links data like CPE, CVE, CWE, CAPEC, MITRE Techniques, and Tactics by providing a bidirectional graph relating these concepts after parsing data from the NVD and MITRE ATT\&CK. Similarly, other researchers~\cite{kiesling_sepses_2019} exploit the capabilities of the Resource Description Framework (RDF) to build a knowledge graph serving the same purpose. These graphs answer questions like: ``\textit{Which Techniques are related to CVE-2022-36804?}''.

Concepts more abstract than CVE can be used to reason in attack graphs. Some researchers developed a tool to visualize how CAPECs could be chained in an information system containing CVEs~\cite{zdemir_snmez_attack_2022}. They manually describe the characteristics and settings of the IS and input some data from the CVSS vector related to the CVEs. Authors of~\cite{falodiya_security_2017} even define a new ontology to describe an attack graph. Nevertheless, they give no methodology about how to fill this ontology, letting think that everything must be hand-made by human experts with good knowledge of the system and its vulnerabilities.

Other studies try to derive the causes and effects of CVEs based on their related description~\cite{urbanska_structuring_2012}. They propose algorithms using keywords and heuristics to extract semantics from the description of the CVE. However, the lack of standardization of CVEs' descriptions complicates the task. Others deal with the problem of CVE enrichment by automatically assigning labels to CVEs based on their descriptions and using natural language processing techniques~\cite{gonzalez_automated_2019}.

%\subsection{Discussion and motivation using an example}
%\label{sec:discussion}
Let us now imagine a simple realistic IS composed of 2 machines, \texttt{m0} and \texttt{m1} where a scan has revealed 3 CVEs. First, the machine \texttt{m0} is running Apache Log4j2 2.0, vulnerable to \texttt{CVE-2021-44228},
and Open Management Infrastructure (OMI) v1.6.8-0, vulnerable to \texttt{CVE-2021-38648}.%(represented Figure~\ref{fig:cve-and-data}).
The NVD database reports that the first CVE enables remote code execution (RCE), while the second allows privilege escalation. Following CVSS, these CVEs have a critical and high severity.
Thirdly, the machine \texttt{m1} hosts \texttt{CVE-2022-36804}, which has a high severity and allows an authenticated RCE.
%To the best of our knowledge, there is no literature on the use of CVEs to design concrete attack scenarios targeting the concerned infrastructure.
To the best of our knowledge, no literature exists on the use of CVEs to design concrete attack scenarios targeting the concerned infrastructure.

\section{\MyFormat{} format definition}
\label{sec:definition}

\begin{figure*}
\begin{lstlisting}[language=json, caption={Generic \MyFormat{}}, label={lst:format}]
["CVE": "CVE-Id",
 "exploit" : "Exploit-URL"
 "vuln_class": one of ("application","operating-system","hardware"),
 "machines_constraints": sublist of ["same","different","same-windows-domain","same-ldap","adjacent-network","unconstrained"],
 "user_source": one of ("application","machine-local","system-or-root","directory","any-user"),
 "user_destination": one of ("application","machine-local","system-or-root","directory"),
 "users_constraints": sublist of ["same","different","same-application"] ]
\end{lstlisting}
    \vspace{-0,5cm}
\end{figure*}

\MyFormat{} relies on the key notion of attack position.
This notion has been introduced in~\cite{poisson_aware, berady:hal-03694719} as a pair $(machine,user)$ designating an attacker who has compromised the account of a user $user$ on a machine $machine$.
%Attack positions can be thought of as nodes in an attack graph where edges depict various ways of spreading itself during an attack.
\MyFormat{} highlights the vulnerability, the source and destination attack position and the particular exploit used to move from the source to the destination attack position.  \MyFormat{} is useful to build attack position graphs, which can represent how the attacker moves laterally and horizontally in an infrastructure.
\MyFormat{} is composed of $7$ elements:
\texttt{CVE}, \texttt{exploit}, \texttt{vuln\_class}, \texttt{machines\_constraints}, \texttt{users\_constraints}, \texttt{user\_source}, \texttt{user\_destination}.
Listing~\ref{lst:format} represent the generic format of \MyFormat{}.
%\vvtt{qq lignes pour expliquer ce qu'il y a dans chacun des paragraphe suivant pour expliquer ce qu'on trouve dans chaque champ l'ensemble des valeurs possibles,
%préciser où on trouve l'info si c'est évident (genre un lien vers la CVE ou l'exploit),
%sinon dire qu'on explique en section suivante comment trouver les valeurs, aides toi si besoin du texte que j'ai commenté ou mieux, maintenant que tu as du recul: réécris. }
%\manu{Quel est le meilleur compromis place/clareté?  Pour gagner de la place on peut fusioner les paragraphes ci-dessous avec le itemize ci-dessus, ils ont exactement les mêmes entrées. Mais on perd un peu en clareté...}
%

\noindent{\colorbox{teal!20!white}{\bf \texttt{CVE}} is the identifier of the form \textit{YEAR-NUMBER} of the CVE.
The same as the one used in the NVD~\cite{nvd_database}.

\noindent{\colorbox{teal!20!white}{\bf\texttt{exploit}} is the url of an exploit applicable to the CVE.
%The different ways to find it are detailed in Section~\ref{sec:fillRunningExploits}

\noindent{\colorbox{teal!20!white}{\bf\texttt{vuln\_class}} specifies the component affected by the vulnerability. Its value can be either  {\tt "application"}, {\tt"operating-system"} or {\tt"hardware"}.

\noindent{\colorbox{teal!20!white}{\bf\texttt{machines\_constraints}} defines the network constraints between the 2 machines involved in the CVE exploitation:
the machine involved in the source attack position, from which the exploit is executed and
the machine involved in the destination attack position, reached after a successful exploitation. The latter is the machine where the CVE is located.
\texttt{machines\_constraints} is a list containing : \texttt{same} when the machines are the same, \texttt{different} when the machines are different, \texttt{unconstrained} when an arbitrary machine can exploit the CVE, \texttt{same-windows-domain} (resp. \texttt{same-ldap}) when the machines must belong to the same Windows domain (resp. same LDAP) and \texttt{adjacent-network} when the machines have to be into two adjacent networks.

% TODO write LDAP with uppercase everywhere

\noindent{\colorbox{teal!20!white}{\bf\texttt{user\_source}} is a string from a predefined set of user account characteristics.
\texttt{user\_source} qualifies the user from which the exploit is executed.
If the user exists only in the vulnerable application (e.g. an account on a website), the value is \texttt{application}.
If it is a local user on a machine, this field's value is \texttt{machine-local}.
If it is a user existing on multiple machines in the same IS, registered in the same LDAP or Active Directory (AD), the value is \texttt{directory}.
Finally, it is \texttt{any-user}, if an arbitrary user can exploit the CVE.

\noindent{\colorbox{teal!20!white}{\bf\texttt{user\_destination}} is a string that represents the characteristics of the user controlled after successful exploitation.
The possible value for this field is \texttt{system-or-root}, \texttt{application}, \texttt{machine-local} or \texttt{directory}.
The semantic of these user's characteristics is the same as the one explained for the values of \texttt{user\_source} even if the characterized user is not necessarily the same.

\noindent{\colorbox{teal!20!white}{\bf\texttt{users\_constraints}} defines constraints linking the user from which the CVE can be exploited (\texttt{user\_source}) and the user accessed after the exploitation (\texttt{user\_destination}).
It is a list containing \texttt{same} when source and destination users are identical, \texttt{different} when they are different and
\texttt{same-application} when source and destination users are accounts on the same application.
For example, \texttt{["different", "same-application"]} indicates that source and destination users are two different user accounts existing one the same application.

    \section{How to populate \MyFormat{}}
\label{sec:fillRunningExploits}
%
%\paragraph*{The methodology}
%In order to ascertain the various properties of a CVE represented in \MyFormat{},
%the pre/post conditions of the CVE in question must be determined.
%\sout{Representation of a CVE using \MyFormat{} can always be done with a deep understanding of the technical details of the CVE in question. However, this level of expertise requires time and does not scale due to the large amount and diversity of CVEs. We propose to gather knowledge about pre/post conditions of a CVE by an easier process that is more generic to every CVEs and more automatable.}

In the following, we detail how to fill \MyFormat{} relative to a \textcolor{applegreen}{\texttt{cve}} and a corresponding exploit \textcolor{applegreen}{\texttt{e}}.

\noindent{\colorbox{teal!20!white}{\bf\texttt{CVE}}  is equal to the identifier of \textcolor{applegreen}{\texttt{cve}}.

\noindent{\colorbox{teal!20!white}{\bf\texttt{Exploit}} is the link towards \textcolor{applegreen}{\texttt{e}}.
For example, an exploit related to a specific CVE-Id can be found in Metasploit using the command \texttt{search cve:} of the msfconsole or in ExploitDB using the search filter of the web interface.
Note that, there is not \MyFormat{} when the exploit is not available.

\noindent{\colorbox{teal!20!white}{\bf\texttt{vuln\_class}} can be filled using the \texttt{part} field (after the second colon in \texttt{CPE 2.3}) of the CPE related to the CVE.
It is \texttt{application, operating-system} or \texttt{hardware} when \texttt{part} is \texttt{a}, \texttt{o} or \texttt{h} respectively.

To populate the other fields requires finding the source attack position ({\tt machine-src}, {\tt user-src}) allowing to execute the exploit \textcolor{applegreen}{\texttt{e}}, and reach the destination attack position.
We propose to do so by first deploying an environment (Docker container, physical or virtual machine) \texttt{machine-dst} vulnerable to \textcolor{applegreen}{\texttt{cve}}. \texttt{machine-dst} is the machine involved in the destination attack position, reached after a successful exploitation of \textcolor{applegreen}{\texttt{cve}}.
Execution of the exploit \textcolor{applegreen}{\texttt{e}} can be tried from different machines and users starting with the ones that are the least constraining for the attacker.

Finding \texttt{machine-src} will allow to fill the field \colorbox{teal!20!white}{\bf\texttt{machine\_constraints}}. It can be done by varying the relationship between the machine from which the exploit is executed and the machine hosting the CVE.
At first, the exploit is executed from an arbitrary machine that is unrelated from the vulnerable environment.
If it succeeds, then \texttt{machines\_constraints} is \texttt{[unconstrained]}.
Else, the exploit's execution can be tried in other contexts.
Deploy 2 machines \texttt{m-src} and \texttt{m-dst} in the same AD domain, make \texttt{m-dst} vulnerable to the CVE and execute the exploit from \texttt{m-src}.
If it succeeds,  then \texttt{machines\_constraints} is \texttt{[different, same-windows-domain]}.
Else, repeat with machines in the same LDAP, or in adjacent networks, or, the most constrained case, execute the exploit from the machine that is hosting the CVE.
Testbeds representing these network configurations with two machines could be prepared.
One machine to execute the exploit and another with a CVE that could vary to get the \MyFormat{} of different CVEs.

The field \colorbox{teal!20!white}{\bf\texttt{user\_source}} can be filled by varying the characteristics of the user who launches the exploit.
First, run the exploit from a user that is unrelated to the vulnerable environment. If it succeeds, then \texttt{user\_source} is \texttt{any-user}.
Else, execute the exploit from a user that is registered in the same AD domain or same LDAP as the vulnerable machine.
If it works, the value is \texttt{directory}.
Else, try the exploitation using a normal local user account, on the vulnerable machine. In case of success, the value is \texttt{machine-local}.
Otherwise, repeat with the user \texttt{root} or \texttt{SYSTEM}. The value is \texttt{system-or-root} if it works.
Finally, create an account on the vulnerable application.
If the exploit succeeds from this account, the value is \texttt{application}.

Once the exploit is fully executed, the goal is to identify the user reachable due to the CVE (i.e. the user involved in the destination attack position) in order to fill the field \texttt{user\_destination}.
In case of an exploit allowing code execution, run the command \texttt{whoami} (existing on UNIX and Windows OS).
If the output is \texttt{root} or \texttt{SYSTEM}, then the value is \texttt{system-or-root}.
Else, run commands like \texttt{Get-AdUser} or \texttt{ldapsearch} to list users in an AD or ldap, when they exist.
If the output of the \texttt{whoami} command is in these lists, then \texttt{user\_destination} is \texttt{directory}.
Else, the user reached is a simple local user and the value is \texttt{machine-local}.
If command execution is not possible, either the exploit leads to an \texttt{application} account or further manual investigation must be made.

The list {\tt users\_constraints} can be filled by comparing \texttt{user\_source}/\texttt{destination} characteristics.
To do so, the command \texttt{whoami} can be run by the user who executed the exploit and by the user reached thanks to the exploit.
If the output is different, then the list of \texttt{user\_constraints} contains \texttt{different}.
It contains \texttt{same}, otherwise.
In the case of \texttt{application} users, who cannot run commands, users of the application affected by the CVE must be listed to see if both source and destination users belong to it. If so,\texttt{same-application} must be added to the list's values.

\begin{figure}[!htpb]
 %   \begin{scriptsize}
 %       \begin{tiny}
\begin{lstlisting}[language=json, breakatwhitespace=true, caption={Three CVE represented in \MyFormat{}}, label={lst:example-cve48}]
{ "CVE": "2021-44228",
"exploit": "https://github.com/rapid7/metasploit- framework/blob/master/modules/exploits/ multi/http/log4shell_header_injection.rb",
"vuln_class": "application",
"machines_constraints": ["unconstrained"],
"users_constraints":  ["different"],
"user_source":"any-user",
"user_destination": "machine-local"},
{ "CVE": "2021-38648",
"exploit": "https://github.com/rapid7/metasploit- framework/blob/master/modules/exploits/ linux/local/cve_2021_38648_omigod.rb"
"vuln_class": "application",
"machines_constraints": ["same"],
"users_constraints":  ["different"],
"user_source":"machine-local",
"user_destination": "system-or-root" },
{ "CVE": "2022-36804",
"exploit": "https://github.com/rapid7/metasploit- framework/blob/master/modules/exploits/ linux/http/bitbucket_git_cmd_injection.rb",
"vuln_class": "application",
"machines_constraints": ["unconstrained"],
"users_constraints":  ["different"],
"user_source":"application",
"user_destination": "machine-local" }
\end{lstlisting}
%    \end{tiny}
%        \end{scriptsize}

\end{figure}

%\manu{Est-ce qu'il ne faudrait pas séparer le reste dans une section à part: exemple des CVE en CAPG et scénario. Aussi, la figure~\ref{fig:attack-scenar-and-graph} pourrai être séparée en 2 : le schéma en Section~\ref{sec:discussion} et le graph à la fin.}
\paragraph{Example 1: \texttt{CVE-2021-44228}}

\begin{figure*}[!htpb]
    \centering
    \begin{subfigure}{0.5\textwidth}
        \includegraphics[width=0.90\textwidth]{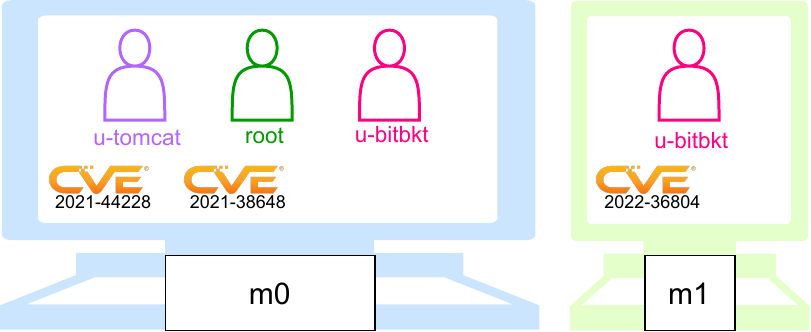}
        \caption{Attack scenario example}
        \label{fig:scenar}
    \end{subfigure}%
    \begin{subfigure}{0.5\textwidth}
        \centering
        \includegraphics[trim=0 0 0 0, clip, width=0.90\textwidth]{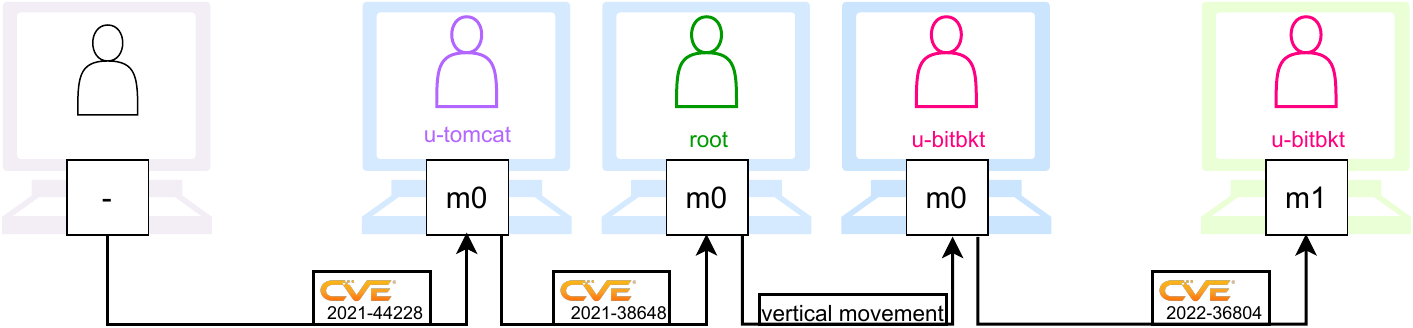}
        \caption{Corresponding attack positions graph}
        \label{fig:graph}
    \end{subfigure}
    \caption{Example of an attack scenario and its corresponding attack positions graph}
    \label{fig:attack-scenar-and-graph}
\end{figure*}

An example of the use of the \MyFormat{} format is depicted in Listing~\ref{lst:example-cve48}.
At first, \texttt{CVE-2021-44228}, which affects Apache Log4j, is represented with a corresponding exploit allowing attackers to execute arbitrary code on a remote machine.
The exploit is accessible on GitHub and can be located using the \textit{search cve:} function in msfconsole. Its url allows to fill the field {\tt exploit}.
From the exploit, we can automatically extract the CVE-Id and pass this parameter to the framework DECRET~\cite{decret} with the command \texttt{python decret.py -n 2021-44228 -r bullseye -s} to automatically deploy a Docker container vulnerable to this CVE.
%A Docker container vulnerable to this CVE have been deployed automatically using the framework DECRET~\cite{decret} with the command \texttt{python decret.py -n 2020-7247 -r bullseye -s}.
%\vvtt{Il faut retravailler la suite de ce texte, à lire ainsi la procédure semble hautement manuelle. Si le paragraph précédent était un peu plus précis, tu pourrais nous dire quel retour de commande permet de conclure à quelle valeur pour chacun des champs }
Then, the exploit can be executed with any users account of any machine by setting the container's IP address as the exploit's target (Exploit's configuration and use can also be scripted.).
Therefore, there are no constraints on the source attack position: \texttt{machines-constraints} is \texttt{unconstrained} and \texttt{user\_source} is \texttt{any-user}.
The exploit execution spawns a shell on the vulnerable environment where the command \texttt{whoami} returns \texttt{tomcat}.
The vulnerable container does not belong to any AD or LDAP, therefore \texttt{tomcat} is a local user: \textit{user\_destination} is \textit{machine-local}.

%As depicted by the two \texttt{constraints} properties of this CVE, the destination attack position involves the same machine and a different user than before the exploitation.
%The user reached is a \texttt{system-or-root}, as it can be seen in the destination attack position (with the id ending with \texttt{eb3}).
%In the case of the attack scenario introduced above, the source attack position would match the control of the user \texttt{u0} on machine \texttt{m0}, the vulnerability host,
%and this CVE could lead to the take-over of the \texttt{system} account on that same machine, later leading to the AD account of the user \texttt{u1}.

\paragraph{Example 2: \texttt{CVE-2021-38648}}
The second CVE in Listing~\ref{lst:example-cve48} represents \texttt{CVE-2021-38648} in \MyFormat{}.
Values of \texttt{user\_source} and \texttt{machine\_constraints} unveil that this CVE can be exploited if an attacker controls a \texttt{machine-local} user on the \texttt{same} machine as the one hosting this CVE.
Successful exploitation leads to a \texttt{different} user than the user in the source attack position: it leads to the user \texttt{root}.
Therefore, the value of \texttt{user\_destination} is \texttt{system-or-root}.

\paragraph{Example 3: \texttt{CVE-2022-36804}}.
The last CVE in Listing~\ref{lst:example-cve48} is \texttt{CVE-2022-36804}.
Its \texttt{user\_source} is \texttt{application} and its \texttt{user\_destination} is \texttt{machine-local}.
This shows that the control of a user account existing on the vulnerable application (i.e. \textit{Bitbucket} for this CVE),
allows to take control of a user existing on the machine hosting the CVE, which could lead to code execution.
\texttt{machines\_constraints} being \texttt{["unconstrained"]} highlights a possible exploitation from an arbitrary machine.

\paragraph{Complete attack scenario}
Let's return to the small example introduced at the end of Section~\ref{sec::relatedWork}. In this small information system, a machine \texttt{m0} hosts 2 CVEs  \texttt{CVE-2021-44228} and \texttt{CVE-2021-38648} and second machine \texttt{m1} host \texttt{CVE-2022-36804} as depicted in Figure~\ref{fig:scenar}.

We also assume {\it at least} the following 3 accounts: \texttt{u-tomcat} is a purely local user of \texttt{m0} able to run Apache Log4j2 2.0, \texttt{root} is a privileged user on \texttt{m0} and
\texttt{u-bitbkt} is the user account of a Bitbucket repository available with the application Server and Data Center 7.0.0, executed on \texttt{m1}.

\noindent{\colorbox{teal!20!white}{\bf Step 1: from any attack position to (\texttt{u-tomcat}, \texttt{m0})} \MyFormat{} relative to \texttt{CVE-2021-44228} specifies that the identified \texttt{exploit} can be successfully used from any machine and with no particular initial user account ({\tt machines\_constraints} is {\tt unconstrained} and {\tt user\_source} is {\tt any-user}).
 This \texttt{exploit} gives to the attacker the control of the local user \texttt{u-tomcat} on machine \texttt{m0} since {\tt user\_destination} is \texttt{machine-local}.

%\vvtt{est ce qu'on parle de l'utilisateur qui en charge de l'exécution de l'application vulnérable (donc appache), je ne pense pas que cela soit clair}\manu{La réponse courte est oui. C'est bien u-tomcat qui exécute apache log4j, mais est-ce que c'est important? Ce qui compte pour le scénario c'est que u-tomcat permet d'exécuter du code sur m0, et sur m0 uniquement}

\noindent{\colorbox{teal!20!white}{\bf Step 2: from (\texttt{u-tomcat}, \texttt{m0}) to (\texttt{root}, \texttt{m0})} \MyFormat{} relative to \texttt{CVE-2021-38648} allows to highlight that the attacker can exploit this vulnerability with the identified  \texttt{exploit} from a local account, in this case \texttt{u-tomcat}, and to take control of the user \texttt{system}.

\noindent{\colorbox{teal!20!white}{\bf Step 3: from (\texttt{root}, \texttt{m0}) to (\texttt{u-bitbkt}, \texttt{m0})} Once the attacker has compromised the  user \texttt{root} he can discover credentials of user \texttt{u-bitbkt}, who can access the Bitbucket repository hosted on \texttt{m1}.

\noindent{\colorbox{teal!20!white}{\bf Step 4: from (\texttt{u-bitbkt}, \texttt{m0}) to (\texttt{u-bitbkt}, \texttt{m1})} The attacker is then able to exploit \texttt{CVE-2022-36804}. \MyFormat{} relative to this last CVE shows that the attacker can run commands on \texttt{m1}.

    \section{Conclusion}
\label{sec::conclu}
Common Vulnerabilities and Exposures (CVEs) are exploited by attackers to progress in a compromised infratructure.
Defensive tools can discover these CVEs but this article points out the lack of precision in the description of CVEs.
Although very useful for measuring the overall security of an information system, knowledge of one or more CVEs does not automatically enable the construction of attack scenarios.

On the other hand, understanding a CVE exploit contains valuable information that can be used to achieve this goal.
 We have proposed here the new format \MyFormat{}, to describe precisely how an attacker could exploit a vulnerability. \MyFormat{} provides operational information to know if a CVE can be exploited and what is the interest of an attacker to exploit it. We also propose and test a methodology for filling in this format, based on the execution of an exploit on a suitable vulnerable environment.  The main purpose of \MyFormat{} is to be used by the defense to highlight critical attack paths that need to be urgently monitored and corrected.

%    \input{apendix}

% \flushend
    \bibliographystyle{IEEEtran}
    \bibliography{mybib}

%\begin{thebibliography}{00}
%\bibitem{b1} G. Eason, B. Noble, and I. N. Sneddon, ``On certain integrals of Lipschitz-Hankel type involving products of Bessel functions,'' Phil. Trans. Roy. Soc. London, vol. A247, pp. 529--551, April 1955.
%\bibitem{b2} J. Clerk Maxwell, A Treatise on Electricity and Magnetism, 3rd ed., vol. 2. Oxford: Clarendon, 1892, pp.68--73.
%\bibitem{b3} I. S. Jacobs and C. P. Bean, ``Fine particles, thin films and exchange anisotropy,'' in Magnetism, vol. III, G. T. Rado and H. Suhl, Eds. New York: Academic, 1963, pp. 271--350.
%\bibitem{b4} K. Elissa, ``Title of paper if known,'' unpublished.
%\bibitem{b5} R. Nicole, ``Title of paper with only first word capitalized,'' J. Name Stand. Abbrev., in press.
%\bibitem{b6} Y. Yorozu, M. Hirano, K. Oka, and Y. Tagawa, ``Electron spectroscopy studies on magneto-optical media and plastic substrate interface,'' IEEE Transl. J. Magn. Japan, vol. 2, pp. 740--741, August 1987 [Digests 9th Annual Conf. Magnetics Japan, p. 301, 1982].
%\bibitem{b7} M. Young, The Technical Writer's Handbook. Mill Valley, CA: University Science, 1989.
%\end{thebibliography}
\end{document}